\documentclass[prl,onecolumn,showpacs,floatfix,amsfonts]{revtex4}
\usepackage{epsfig,graphicx,times}
\usepackage{amstext}
\usepackage{amsmath}
\usepackage{amssymb}
\usepackage{graphicx}
\usepackage{latexsym}
\usepackage{bm}
\usepackage{color}
\def\dbarit {{\mathchar'26\mkern-11mud}}

\def \dbarit {{\mathchar'26\mkern-11mud}}

\begin{document}

\title{Maximum Efficiency of Ideal Heat Engines Based on a Small System: Correction to the Carnot Efficiency at the Nanoscale}
\author{H. T. Quan}
\affiliation{School of Physics, Peking University, Beijing 100871, China}
\affiliation{Collaborative Innovation Center of Quantum Matter, Beijing 100871, China}

\begin{abstract}
We study the maximum efficiency of a heat engine based
on a small system. It is revealed that due to the finiteness of the
system, irreversibility may arise when the working substance contacts with a heat reservoir.
As a result, there is a working-substance-dependent
correction to the Carnot efficiency. 
We derive a general and simple expression for the maximum efficiency of a
Carnot cycle heat engine in terms of the relative entropy. This maximum
efficiency approaches the Carnot efficiency asymptotically when the size of the working substance increases to the thermodynamic limit.
Our study extends Carnot's result of the maximum efficiency to an arbitrary working substance and demonstrates the subtlety of thermodynamic laws in small systems.
\end{abstract}

\maketitle

\emph{Introduction:} Carnot conceived an ideal thermodynamic cycle,
which consists of two isothermal and two adiabatic processes \cite{carnot}. This
cycle, now known as Carnot cycle, has the highest efficiency among all
thermodynamic cycles. The Carnot cycle is of great importance in
the development of the principles of thermodynamics, especially the second
law of thermodynamics. 
For a Carnot cycle, its efficiency is given by $\eta_{C}=1-\frac{T_{C}}{T_{H}}$, where
$T_{C}$ and $T_{H}$ describe the temperatures of the cold and the hot heat
reservoirs respectively. Since Carnot's era it has been widely accepted that the maximum efficiency (the Carnot efficiency) 
does not depend on the details of the working substance~\cite{carnot} (when the working substance is in the thermodynamic limit, the Carnot efficiency is synonymous to the maximum efficiency of a heat engine).
In most textbooks the Carnot efficiency is derived by utilizing thermodynamic methods, such as using the thermodynamic (or absolute) temperature scale (see Ref~\cite{zemansk} for example), in which it was assumed implicitly that the working substance is in the thermodynamic limit.
One exception, however, is Gibbs' book \cite{gibbs}, where the Carnot efficiency is derived from the first principle.
In obtaining the result, it has been assumed implicitly that
the system is always in a canonical distribution in all the four thermodynamic processes \cite{tolmann}. In the isothermal processes, this is certainly true.
In the adiabatic processes, however, this assumption is not obviously consistent with classical mechanics \cite{ken00} (or quantum mechanics \cite{quan07}), because Hamiltonian dynamics (or quantum mechanics) does not necessarily maintain the canonical distribution.
The subtlety is that for the working substance in the thermodynamic limit, the above assumption is always valid. Nevertheless, for the working substance in the opposite limit, i.e., for a small system
consisting of a few molecules, this assumption usually does not hold true \cite{ken00} and it may lead to a non-negligible correction to the Carnot efficiency \cite{ken00} (see also \cite{linden10}), or alternatively, in a small system, the Carnot efficiency is no longer synonymous to the maximum efficiency of an ideal heat engine.

In recent years, thermodynamics of
small systems has attracted a lot of attention \cite{smallsystem, horokecki}.
Some theorems concerning far from
equilibrium processes have been discovered \cite{fluctuation, JE1, crooks}, and have been verified experimentally in small systems, for example, in a single RNA molecular chain \cite{experiment}. In the field of thermodynamics in small systems an important question is the validity of the thermodynamic laws when the system is in the opposite regime of the thermodynamic limit - only a few particles are involved.
In recent years some investigations have been carried out to reexamine the validity of the laws and principles
of thermodynamics in small systems. For
example in Ref. \cite{ken02,tasaki}, it has been pointed out that
irreversibility may arise when the system contacts with a heat reservoir due to the
finiteness of the system. In Ref. \cite{ken00} it has been pointed out
that the Carnot efficiency may not be achieved in a small system. In the
study of its quantum analogue, it has been emphasized in Refs.
\cite{quan07} (see also \cite{ken00,ken02,arnaud}) that two conditions (i) the process must be
quasi-static and (ii) all energy levels change in the same ratio in the adiabatic process
must be satisfied in order to achieve the Carnot efficiency. When
either condition is not satisfied, the maximum efficiency of the
heat engine will be lower than the Carnot efficiency. Despite of the
rapid progress in this research field in the last few years, a key question remains to be answered:
What is the maximum efficiency of an ideal heat engine when the working substance is
much less than the thermodynamic limit? and if the maximum efficiency is lower than the Carnot efficiency, how does the result reconcile with the Carnot's theorem in the thermodynamic limit? 

In this letter we address this question and 
derive a simple and general expression for the
maximum efficiency of an ideal heat engine 
based on an arbitrary system. 
For macroscopic system, the maximum efficiency reproduces the Carnot efficiency. For a microscopic system,
the maximum efficiency approaches the Carnot efficiency only in special cases, i.e., when the all energy levels of the system change in the same ratio in the adiabatic process. Otherwise a correction to the Carnot efficiency occurs.
We would like to mention that the efficiency of a Carnot cycle at the maximum
power \cite{maximumpower1} has been extensively studied in recent years, but our
emphasis is different. We only consider the quasi-static process or
zero power case. Instead of focusing on the maximum power, our
emphasis is put on the correction to the Carnot efficiency due to the
finite size effect. For simplicity, we will focus on the classical system. The extension of our current study to
quantum systems \cite{kosloff} is straightforward. 

\begin{figure}[tbh]
\includegraphics[bb=66 331 517 685, width=8cm, clip]{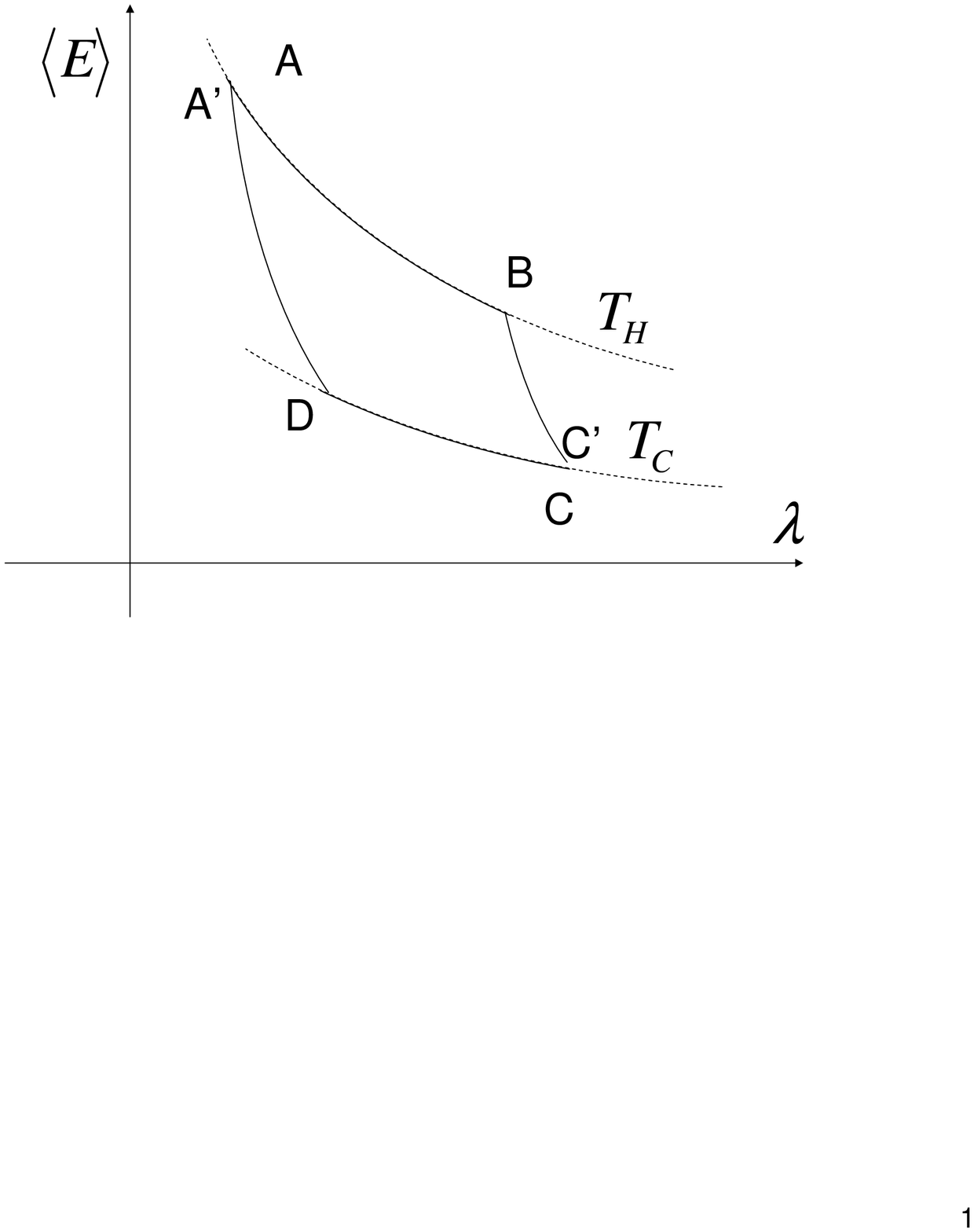}
\caption{ Schematic diagram of a Carnot cycle. The two axes
represent the average energy $\left \langle E\right \rangle$ of the system and a controlling parameter $\lambda$ respectively. $A\to B$ represents an isothermal process with the heat reservoir at
temperature $T_{H}$. $C\to D$ represents another isothermal process with the heat reservoir at temperature $T_{C}$. $B\to C'$ and $D \to A'$ represent
two adiabatic processes. All these processes proceed quasistatically. When the distribution of the system at $C'$ ($A'$) is not canonical, there is an irreversible relaxation process of the system from $C'$ ($A'$) to $C$ ($A$). } \label{scheme}
\end{figure}

\emph{Carnot cycle at the nanoscale:} As is known a Carnot cycle consists of two
isothermal and two adiabatic processes.  When all the processes
proceed quasi-statically or reversibly, the efficiency approaches
its maximum value, $\eta=1-\frac{T_{C}}{T_{H}}$ which depends
only on the temperatures of the two heat reservoirs, and has nothing
to do with the details of the working substance \cite{carnot}. It has been assumed
that this efficiency can accurately describe the maximum efficiency
of an ideal heat engine as long as all four processes are quasi-static. For a system
with a huge number of degrees of freedom, e.g., a system consists of $10^{23}$ molecules, this is obviously true. However, when the
system is small, i.e., a few molecules are involved, special attention need be paid to the applicability of the principles of
thermodynamics  \cite{ken00,ken02}.
Let us recall that a Carnot cycle consists of four processes (See Fig. \ref{scheme}). It
starts from an isothermal process ($A \to B$). In order to achieve
the maximum efficiency, this process proceeds quasi-statically, so
that the system is always in a canonical distribution at every instant of the
process and its temperature is always equal to the temperature of
the heat reservoir $T_{H}$. After this process, the system of interest is
taken out of the heat reservoir at instant $B$. Some
external parameters are manipulated quasistatically and adiabatically in the following
process ($B \to C'$). This is an adiabatic process. In order to achieve the maximum efficiency, this process also needs to proceed slow enough, or quasistatically. At the end of this adiabatic process ($B \to C'$), the system may or may not be in a
canonical distribution even though at the beginning of the process (at instant $B$) the system is. Hence, in
general one cannot use an effective temperature to describe the
system after the adiabatic process \cite{ken00,ken02} unless the system satisfies certain conditions or its size is large enough (We will clarify later that for a system in the thermodynamic limit, it is very close to a canonical distribution at instant $C'$, so that one can
still assign an effective temperature to it). When the system is not in a canonical distribution
after the adiabatic process, there is an irreversible relaxation process of
the system soon after it is put into contact with a heat reservoir (we have
carefully chosen the parameter $\lambda_{C}$ such that the
average energy of the system does not change after the relaxation
process.) \cite{ken02}. We use $C'$ and $C$ to denote the states of
the system before and after the relaxation. After the relaxation the
system reaches a canonical distribution and the temperature of the
system is equal to that of the lower temperature heat reservoir. 
We will see that in the thermodynamic limit, the effect of the relaxation process
can be ignored, because the entropy increases in the relaxation process is negligibly small in comparison with the entropy of the system. 
Following the relaxation process there is another isothermal compression process ($C \to D$). This is
similar to the isothermal expansion process in which the system is
always in equilibrium with the lower temperature heat reservoir (it is always
in a canonical distribution during this process). After the isothermal
compression process another adiabatic compression process ($D \to A'$)
follows. Similar to the adiabatic expansion process ($B \to C'$), the small
system may not be in a canonical distribution during the adiabatic
process even though it proceeds quasi-statically. In order to
restore the working substance to its original state, we need to put the
system into contact with the heat reservoir at temperature $T_{H}$ again. The
process is irreversible, because the system will relax from a
non-canonical distribution $A'$ to a canonical one $A$. In the relaxation process, although the average
energy does not change, the statistical entropy of the ensemble increases
in this relaxation process. The reason is that for a given average energy, the distribution which maximizes the entropy is the canonical distribution.

\begin{figure}[tbh]
\includegraphics[width=6cm, angle=270, clip]{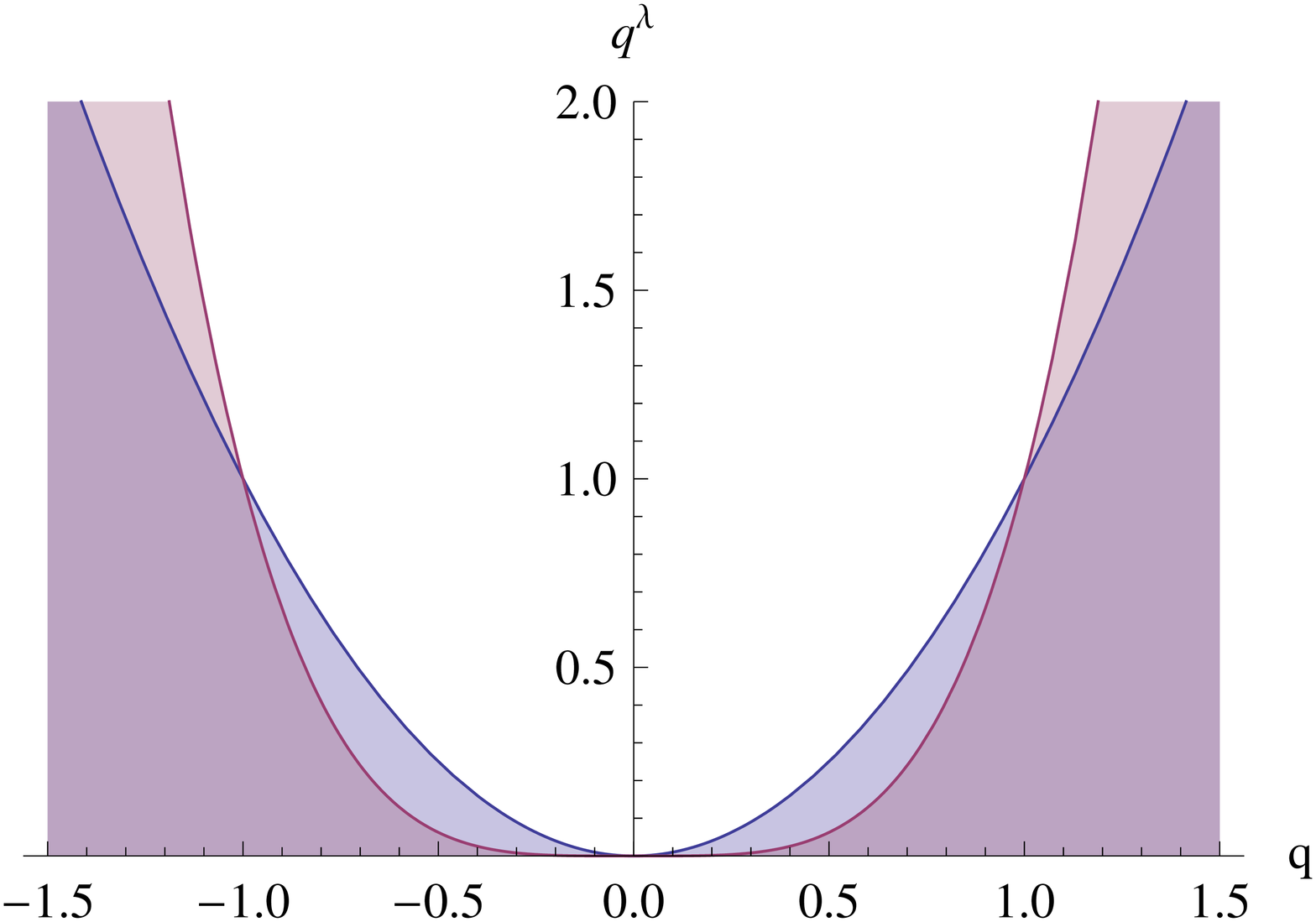}
\caption{ Schematic diagram of the potential well $|a q|^{\lambda(t)}$. Here we have chosen $a=1$, $\lambda_{B}=2$ and $\lambda_{D}=4$.
In the adiabatic process $B\to C'$ when the potential well is varied, the system does not obey the canonical distribution (at a varying temperature) if it starts from a canonical distribution.} \label{figpotential}
\end{figure}

In order to describe the whole thermodynamic cycle, we use a
parameter $\lambda$ to describe the cycle (See Fig. \ref{scheme}). The thermodynamic cycle
can be completely determined by four parameters $T_{H}$, $T_{C}$, $\lambda_{B}$, and $\lambda_{D}$.
The other two parameters $\lambda_{A}(=\lambda_{A'})$ and
$\lambda_{C}=(\lambda_{C'})$ can be determined
by the conditions of the average energy during the relaxation process
\begin{equation}
\begin{split}
\left \langle E_{l} \right \rangle &\equiv\int d\Gamma_{l}\rho^{\mathrm{eq}}(\lambda_{l}, \Gamma_{l}) H(\lambda_{l}, \Gamma_{l})\\
&= \int d\Gamma_{l'}\rho(\lambda_{l'}, \Gamma_{l'}) H(\lambda_{l'}, \Gamma_{l'})\equiv\left \langle E_{l'} \right \rangle , (l=A,C).
\end{split}
\label{conservation}
\end{equation}
We consider $l=C$ as an example. Here $\Gamma_{C}$ represents a point in the phase space with the parameter $\lambda_{C}$;
$\rho^{\mathrm{eq}}(\lambda_{C}, \Gamma_{C}) \propto \exp{[-\beta_{C} H(\lambda_{C}, \Gamma_{C})]}$ describes the equilibrium (canonical)
distribution of the microscopic states in the phase space with the parameter $\lambda_{C}$ and the temperature $T_{C}$ (point $C$ in Fig. \ref{scheme}); $H(\lambda_{C}, \Gamma_{C})$
describes the energy of the system; $\rho(\lambda_{C'}, \Gamma_{C'})$ describes the density distribution of the states of the system in the phase space evolving from an
equilibrium (canonical) distribution $\rho^{\mathrm{eq}}( \lambda_{B}, \Gamma_{B}) \propto \exp{[-\beta_{H} H(\lambda_{B}, \Gamma_{B})]}$, and $\rho(\lambda_{C'}, \Gamma_{C'})$
is not necessarily to be in an equilibrium (canonical) distribution \cite{ken00}.
The heat absorbed $\dbarit Q_{H}$ in the isothermal expansion process ($A \to B$) and released $\dbarit Q_{C}$ in the isothermal contraction process ($C \to D$) can be expressed as
\begin{equation}
\begin{split}
\dbarit Q_{H}=T_{H} (S_{B}-S_{A}),\\
\dbarit Q_{C}=T_{C} (S_{C}-S_{D}),
\end{split}\label{heat}
\end{equation}
where $S_{i}=-k_{B} \int d\Gamma_{i} \rho^{\mathrm{eq}}(\lambda_{i}, \Gamma_{i}) \ln \rho^{\mathrm{eq}}( \lambda_{i}, \Gamma_{i})$,
$(i=A, B, C, D)$ is the equilibrium thermodynamic entropy; $k_{B}$ is the Boltzmann's constant. The efficiency of the thermodynamic cycle can be expressed as the $\eta_{\mathrm{max}}\equiv\frac{\dbarit W}{\dbarit
Q_{H}}=1-\frac{\dbarit Q_{C}}{\dbarit Q_{H}}$. Substituting Eq. (\ref{heat}) into
the expression of the efficiency we obtain
\begin{equation}
\eta_{\mathrm{max}}=1-\frac{T_{C} [(S_{C'}-S_{D})+ (S_{C}-S_{C'})]}{T_{H}
[(S_{B}-S_{A'})-(S_{A}-S_{A'})]}, \label{Carnot}
\end{equation}
where $S_{j}=-k_{B} \int d\Gamma_{j} \rho(\lambda_{j}, \Gamma_{j}) \ln \rho( \lambda_{j}, \Gamma_{j})$,
$(j=A', C')$ is the statistical entropy of the ensemble after the adiabatic evolution $D \to A'$ and $B \to C'$.
Because of the Liouville theorem, the statistical entropy of the
working substance remains a constant in the adiabatic processes
$S_{B}=S_{C'}$, $S_{D}=S_{A'}$.
In addition, the change of the entropy of the
system in the relaxation process can be expressed as a relative
entropy \cite{ken02} due to the conditions of the average energy (\ref{conservation})
\begin{equation*}
\begin{split}
S_{l}-S_{l'}&=D[ \rho_{l'} ||\rho_{l}^{\mathrm{eq}} ]\\
&\equiv \int d \Gamma_{l'} \rho(\lambda_{l'}, \Gamma_{l'}) \ln \frac{\rho(\lambda_{l'}, \Gamma_{l'})}{\rho^{\mathrm{eq}}(\lambda_{l}, \Gamma_{l'}) }, (l=A, C),\\
\end{split}
\end{equation*}
where $D[\rho_{l'}  || \rho_{l}^{\mathrm{eq}}]$, $(l=A, C)$ is the relative entropy \cite{relativeentropy1} of the two distributions of ensembles in the phase space. Relative entropy is a widely used quantity in the studies of (quantum) information theory \cite{relativeentropy2} and non-equilibrium thermodynamics \cite{ken02,kawai,ueda11,polkovnikov}. This quantity is
non-negative, and is equal to zero only when the two distributions are identical $\rho_{l}^{\mathrm{eq}} \equiv \rho_{l'}$ \cite{nonnegative}, or in our case the system changes from
one canonical distribution to another.
Substituting the above equation into Eq. (\ref{Carnot}), the maximum efficiency can be expressed in terms of the relative entropy \cite{explain}
\begin{equation}
\eta_{\mathrm{max}}=1-\frac{T_{\mathrm{C}}\{(S_{B}-S_{D})+D[ \rho_{C'}||\rho_{C}^{\mathrm{eq}} ]\}}{T_{\mathrm{H}}
\{(S_{B}-S_{D})-D[ \rho_{A'} || \rho_{A}^{\mathrm{eq}}]\}}.\label{efficiency}
\end{equation}
This is the main result of our letter. Because of the facts $S_{B}-S_{D}>0$ and both relative entropies $D[ \rho_{C'} || \rho_{C}^{\mathrm{eq}}]$
and $D[\rho_{A'} || \rho_{A}^{\mathrm{eq}} ]$ are nonnegative \cite{nonnegative}, from the expression of the efficiency
(\ref{efficiency}) we immediately obtain the relation $\eta_{\mathrm{max}} \leq
\eta_{C}=1-\frac{T_{\mathrm{C}}}{T_{\mathrm{H}}}$. Here, the equality holds
only when both $D[  \rho_{C'}|| \rho_{C}^{\mathrm{eq}}]$ and $D[\rho_{A'} || \rho_{A}^{\mathrm{eq}} ]$ are equal to zero. We
emphasize that the efficiency obtained here is universal, because it 
holds irrespective of the working substance. From the Carnot
theorem, one would naturally expect
that when the working substance is in the thermodynamic limit $N\to \infty$ (here $N$ is the number of the particles of the system) the above maximum efficiency (\ref{efficiency})
approaches the Carnot efficiency so that our result can reconcile with Carnot's theorem. That is, in the thermodynamic limit, the relative entropy
$D[\rho_{A'} || \rho_{A}^{\mathrm{eq}} ]$ and $D[ \rho_{C'}|| \rho_{C}^{\mathrm{eq}} ]$ should be negligibly small in comparison with $S_{B}-S_{D}$.
Because both $S_{B}$ and $S_{D}$ are thermodynamic entropy, and hence are extensive quantities $S_{B} \propto N$, $S_{D} \propto N$, we expect that in the thermodynamic limit, the relative entropy increases slower than $N$:
\begin{equation}
\lim_{N\to \infty} \frac{D[ \rho_{l'} || \rho_{l}^{\mathrm{eq}}]}{N} \to 0, (l=A, C).\label{vanishing}
\end{equation}
A rigorous proof of this result is given in the supplemental material. In the following we use a simple example to demonstrate our main result. We will show that in the extreme limit of a few particles the correction is non-negligible. The correction to the Carnot efficiency decreases with the increase of the number of the particles involved.
In the thermodynamic limit, our result reproduces the Carnot efficiency.
Hence, our result of the maximum efficiency (equation (\ref{efficiency})) includes the Carnot efficiency as a special case. 

\begin{figure}[tbh]
\includegraphics[width=8cm, clip]{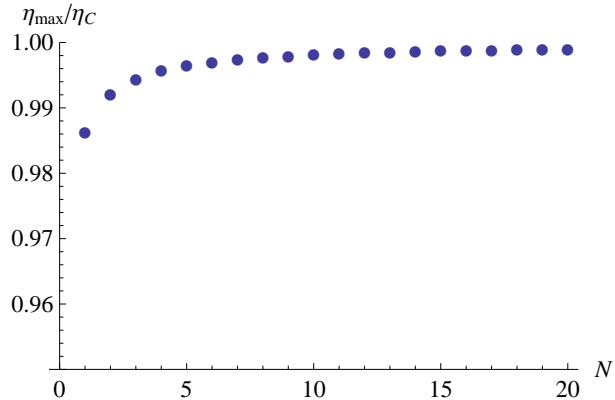}
\caption{Relative efficiency of a Carnot cycle as a function of the particle number $N$. Here we choose the Boltzmann constant $k_{B}=1$, the temperatures $T_{H}=4.6$, $T_{C}=2.3$, and the parameters of the cycle $\lambda_{B}=2$, and $\lambda_{D}=4$.} \label{relativeefficiency}
\end{figure}

\emph{Example:} In order to achieve the Carnot efficiency, the key
requirement is that the system always be
in a canonical distribution (at a varying temperature) in the two adiabatic processes $B \to C'$ and $D \to A'$ \cite{ken00,quan07,arnaud}. 
For classical systems this is equivalent to the requirement that the energy of the system changes in the
same ratio in the adiabatic process. Let us consider
a system consisting of $N$ weakly coupled particles described by the following Hamiltonian
\begin{equation}
H=\sum_{\alpha=x,y,z}\sum_{i=1}^{N}\frac{p_{i, \alpha}^{2}}{2m}+\left |a(t) q_{i, \alpha} \right|^{\lambda(t)}+V,\label{potential}
\end{equation}
where $m$ is the mass of a particle; $a(t)$ and $\lambda(t)$ are two controlling parameters; $q_{i, \alpha}$ and $p_{i, \alpha}$ denote the position and the moment of the $\alpha$ ($\alpha=x$, $y$, $z$) degree of freedom of the $i$th particle. The interactions among these particles $V$ are so weak that they can be
ignored in comparison with the kinetic and potential energy of the particles, but they are strong enough to make the $N$-particle system to be ergodic. When one fixes
$\lambda=2$ and varies $a(t)$, this is a forced harmonic oscillator. 
It can be easily checked that when $\lambda$ is fixed, and $a(t)$ is varied quasistatically and adiabatically, the system is
always in a canonical distribution (at a varying effective temperature) as long as one starts from a canonical
distribution \cite{ken02,quan07}, because the energies of different microscopic states change in the same ratio in this adiabatic process \cite{chris07}.
 However, if one fixes $a$ and varies $\lambda(t)$ quasistatically and adiabatically (Fig. \ref{figpotential}),
 the system will not obey the canonical distribution in the adiabatic process if it
starts from a canonical distribution, because the energies of different microscopic states do
not change in the same ratio. As an example, we fix $a$, and vary the exponent $\lambda(t)$
from instant $B$ ($\lambda_{B}=2$, harmonic oscillator potential) to instant $C'$
($\lambda_{C}$) quasi-statically (Fig. \ref{figpotential}). 
We calculate the relative entropy $D[  \rho_{C'}|| \rho_{C}^{\mathrm{eq}}]$ and study how it changes with the particle
number $N$. We find that it reaches an asymptotic value in the large $N$ limit. The procedure (see the supplemental material) of calculating the relative entropy can be summarized as follows:
For $N$ weakly interacting particles described by the Hamiltonian (\ref{potential}), one can calculate the volume of the phase space as a function of the energy of the energy shell. 
At the end of the adiabatic process, at instant $C'$, $\lambda=\lambda_{C}$, the energy of the $N$ interacting
particle system can be determined by the adiabatic invariant \cite{ken00,ken02,adiabaticinvariance,chris07}.
By using Eq. (\ref{conservation}) one can determine $\lambda_{C}$. Thus the thermodynamic entropy $S_{C}$ can be
calculated. Also, due to the Liouville theorem,
the entropy before and after the adiabatic process should be equal $S_{C'}=S_{B}$. Hence one obtains the increase of the entropy
in the relaxation process $S_{C}-S_{C'}$, which is equal to the relative entropy $D[  \rho_{C'}|| \rho_{C}^{\mathrm{eq}}]$.
Through a detailed calculation (see the supplemental material)
we find
\begin{equation}
\begin{split}
&D[\rho_{C'}||\rho^{eq}_{C}]=3Nk_{B}\frac{2-\lambda_{C}}{2 \lambda_{C}}+ k_{B}\ln{\left[ \frac{\Gamma\left(3N \frac{2+\lambda_{C}}{2 \lambda_{C}}+1\right)}{\Gamma\left(3N+1\right)} \right]}+ \\
&3Nk_{B}\frac{2+\lambda_{C}}{2 \lambda_{C}}
\left\{\ln {\frac{\Gamma\left(3N+\frac{2\lambda_{C}}{(2+\lambda_{C})}\right)}{\Gamma\left(3N\right)} }
- \ln \left[ 3N \frac{2+\lambda_{C}}{2 \lambda_{C}} \right] \right\},\\
\end{split}
\label{correction}
\end{equation}
where $\Gamma(x)$ is the Gamma function. 
Similarly we obtain an expression for $D[ \rho_{A'} || \rho_{A}^{\mathrm{eq}} ]$.
Substituting them into Eq.~(\ref{efficiency}) we obtain the maximum efficiency and the correction to the
Carnot efficiency (see Fig. \ref{relativeefficiency}). 
Meanwhile, from Eq.~(\ref{correction}) one can see that $\lim_{N\to \infty} D[  \rho_{l'}|| \rho_{l}^{\mathrm{eq}}]/N \to 0$ is satisfied.
 This means when the particle number is large, not necessarily to be close to $N \sim
10^{23}$, the correction is already vanishingly small in comparison with
$S_{B}-S_{D}$, which is proportional to $N$. Hence, in the thermodynamic limit the maximum efficiency reproduces the Carnot efficiency $\eta_{\mathrm{max}}\to\eta_{C}=1-\frac{T_{C}}{T_{H}}$.
Nevertheless, when the particle number is in the opposite
limit $N\to 1$, the correction to the Carnot efficiency may have observable consequence in experiments.
In Fig. \ref{relativeefficiency} we plot the maximum efficiency as a function of the particle number. It can be seen that this correction to the Carnot efficiency is less prominent when the particle number is larger than 20. Although the deviation from the Carnot efficiency in our example is tiny (less than two percent), it
well serves the purpose to demonstrate our main results.

One can understand the correction to the Carnot efficiency through
the following fact: the canonical ensemble and the
microcanonical ensemble are equivalent in the thermodynamic limit.
However, these two ensembles are not equivalent when the system is in the extremely small limit.


\emph{Discussion and conclusion:}
The Carnot efficiency is a milestone in the development of the principles of traditional thermodynamics.
Since Carnot's seminal work \cite{carnot} in 1820s it has been widely accepted that the maximum efficiency of an
ideal heat engine depends only on the temperatures of the two heat reservoirs, and does not depend on the details
of the working substance. In this letter we reexamine the maximum efficiency of a heat engine based on a small system.
We find, however, this statement may not hold true
when the working substance consists of only a few particles (at the nanoscale). The Carnot efficiency is not synonymous to the maximum efficiency of an ideal heat engine based on a small system if the variation of the potential is fixed. 
By utilizing the Liouville theorem and the adiabatic invariant, we derive a universal
and simple expression for the maximum efficiency of a heat engine (\ref{efficiency}).
Our study reveals that irreversibility may arise due to the finiteness of the working substance.
As a result a correction to the Carnot efficiency arises. Different from the usual belief that
the maximum efficiency is independent of the details of the working substance, we find a working-substance-dependent correction to the
Carnot efficiency, which is expressed in terms of the relative entropy.
This result is valid for both small systems and large systems. 
The correction to the Carnot efficiency decreases with the increase of the system size (particle number) and vanishes for systems in the thermodynamics limit. In the large system limit, our result reproduces the Carnot efficiency. 
In this sense our result can be regarded as a generalization of the Carnot's theorem, and our investigation elucidates one subtlety of the second law of thermodynamics in small systems. Hopefully, the theoretical predictions presented here can be
tested by employing the cutting-edge techniques developed in the field of ultra-cold atoms. Proposals for experimental verification of our theoretical predictions will be given later.

{\bf Acknowledgments:} The author thanks C. Jarzynski for helpful discussions and gratefully acknowledges the support from the
National Science Foundation of China, grant 11375012, and The Recruitment Program of Global Youth Experts of China.

\section{supplemental material}

\subsection{A: PROOF OF EQUATION (5) OF THE MAIN BODY}
Equation (5) in the main body is
\begin{equation}
\lim_{N\to \infty} \frac{D[ \rho_{l'} || \rho_{l}^{\mathrm{eq}}]}{N} \to 0, (l=A, C).
\end{equation}
We will prove the case of $l=C$. The case of $l=A$ can be proved in a similar way. The sketch of the proof of this equation can be outlined as follows: The system consists of $N$ particles. At instant $B$, the canonical ensemble $\rho_{B}^{\mathrm{eq}}$ is indistinguishable from a microcanonical ensemble of the $N$-particle system when $N$ approaches infinity $(N \to \infty)$. After the quasistatic process, at instant $C'$, the ensemble $\rho_{C'} $ is also very close to a microcanonical ensemble if $N$ approaches infinity $(N \to \infty)$. Let us consider the subsystem of a single particle, and denote its density distribution function with $\rho_{\mathrm{sub}}$. At point $C'$, the subsystem is exactly a canonical ensemble if $N \to \infty$
\begin{equation}
\lim_{N\to \infty} D[ \rho_{\mathrm{sub}} || \rho_{\mathrm{sub}}^{\mathrm{eq}}] \to 0\label{sub},
\end{equation}
where $\rho_{\mathrm{sub}}^{\mathrm{eq}}$ is the equilibrium (canonical) distribution function of a single particle. It can be obtained from the equilibrium density distribution function of $N$ particle $\rho_{C}^{\mathrm{eq}}$. For a system of $N$ particles (weakly coupled and decomposable), we have
\begin{equation}
 D[ \rho_{C'} || \rho_{C}^{\mathrm{eq}}]=N D[ \rho_{\mathrm{sub}} || \rho_{\mathrm{sub}}^{\mathrm{eq}}].\label{proportion}
\end{equation}
Combining Eq.~(\ref{sub}) and Eq.~(\ref{proportion}) we obtain the following relation
\begin{equation}
\lim_{N\to \infty} \frac{D[ \rho_{C'} || \rho_{C}^{\mathrm{eq}}]}{N} \to 0.
\end{equation}

\subsection{B: CALCULATION OF THE RELATIVE ENTROPY}

{\bf Number of states in the phase space enclosed by the energy shell for a $N$-particle system in a $d$-dimensional potential:}
Let us consider a system consisting of $N$ particles in $d$ dimension. The Hamiltonian of the system can be written as
\begin{equation}
H=\sum_{\alpha=1}^{d}\sum_{i=1}^{N}\frac{p_{i, \alpha}^{2}}{2m}+\left |a q_{i, \alpha} \right|^{\lambda(t)}+V.
\end{equation}
Here $q_{i, \alpha}$ and $p_{i, \alpha}$ denote the position and the moment of the $\alpha$th degree of freedom of the $i$th particle. The interactions among particles $V$ are so weak that they can be
ignored in comparison with the kinetic and potential energy of the particles. Meanwhile they are strong enough to ensure the $N$-particle system to be ergodic. In the following we will derive Eq. (7) of the main body.
Let us start from calculating the volume of the phase space as a function of the total energy $E$. 
The partition function of the $N$ particle system is
\begin{equation}
\begin{split}
Z_{N}=&\frac{1}{N!}\left[ \frac{1}{h}\int^{+\infty}_{-\infty} dp \int^{+\infty}_{-\infty} dq \exp {\left[-\beta \left(\frac{p^{2}}{2m}+\left| a q \right|^{\lambda}\right)\right]} \right]^{dN}\\
=&\frac{1}{N!}\left[ \frac{1}{h} \sqrt{2 \pi m} \frac{2}{a}\Gamma \left(1+\frac{1}{\lambda}\right) \right]^{dN} \beta^{-\frac{dN}{\lambda}-\frac{dN}{2}},
\end{split}
\end{equation}
where $h$ is Planck's constant, $\Gamma(x)$ is the Gamma function and we have used the following relation
\begin{equation}
\int_{0}^{\infty} dx \exp{\left[ -ax^{b}\right]}=a^{-\frac{1}{b}}\Gamma \left(1+ \frac{1}{b}\right).
\end{equation}
The density of states of the $N$ particle system
\begin{equation}
\begin{split}
\frac{\partial \Phi(E)}{\partial E} = g(E)&=\frac{1}{2\pi i} \int^{\beta'+i\infty}_{\beta'-i\infty} e^{\beta E} Z_{N} d\beta, (\beta'>0),\\
Z_{N}&=\int_{0}^{\infty}g(E) e^{-\beta E} dE,
\end{split}
\end{equation}
where $\Phi(E)$ is the number of states in the phase space enclosed by the energy shell characterized by the total energy $E$. The density of state for the $N$ particle system in a power law potential is
\begin{equation}
g(E)=\frac{1}{2\pi i} \int^{\beta'+i\infty}_{\beta'-i\infty} e^{\beta E}\frac{1}{N!}\left[ \frac{1}{h} \sqrt{2 \pi m} \frac{2}{a}\Gamma \left(1+\frac{1}{\lambda}\right) \right]^{dN} \beta^{-\frac{dN}{\lambda}-\frac{dN}{2}} d\beta.
\end{equation}
By utilizing the relation
\begin{equation}
\frac{1}{2\pi i} \int_{s'-i \infty}^{s'+ i \infty} \frac{\exp{(sx)}}{s^{n+1}}ds=\left\{ \begin{array}{c}
\begin{split}
&\frac{x^{n}}{n!}, (x>0)  \\
&0, (x<0)
\end{split}
\end{array}
\right.,
\end{equation}
we have
\begin{equation}
\begin{split}
g(E)=&\frac{1}{N!}\left[ \frac{1}{h} \sqrt{2 \pi m} \frac{2}{a}\Gamma \left(1+\frac{1}{\lambda}\right) \right]^{dN} \frac{1}{2\pi i} \int^{\beta'+i\infty}_{\beta'-i\infty} e^{\beta E} \beta^{-\frac{dN}{2}-\frac{dN}{\lambda}} d\beta\\
=&\frac{1}{N!}\left[ \frac{1}{h} \sqrt{2 \pi m} \frac{2}{a}\Gamma \left(1+\frac{1}{\lambda}\right) \right]^{dN} \frac{E^{\frac{dN}{2}+\frac{dN}{\lambda}-1}}{\Gamma\left(\frac{dN}{2}+\frac{dN}{\lambda}\right)}.
\end{split}
\end{equation}
The number of states in the phase space enclosed by the energy shell is then given by
\begin{equation}
\begin{split}
\Phi(E,\beta)=&\int^{E}_{0}g(E')dE'\\
=&\frac{1}{N!}\left[\frac{1}{h} \sqrt{2 \pi m} \frac{2}{a}\Gamma \left(1+\frac{1}{\lambda}\right) \right]^{dN}  \frac{E^{\frac{dN}{2}+\frac{dN}{\lambda}}}{\Gamma\left(\frac{dN}{2}+\frac{dN}{\lambda}+1\right)}.
\end{split}
\end{equation}

{\bf Determining parameters of the thermodynamic cycle of the heat engine:}
We assume the particles are weakly coupled, so the $N$-particle system is ergodic and the adiabatic invariant of the system can be used to determine the final energy of the system
from its initial energy. The system is initially prepared in a microcanonical ensemble with energy $E=E_{0}$ and the potential is characterized by $\lambda=\lambda(0)$. When the parameter is adiabatically ramped to $\lambda(\tau)$, the ensemble remains in a microcanonical ensemble due to the ergodicity of the system. The final energy of the system can be calculated from the adiabatic invariant $\Phi[\lambda(0),E_{0}]=\Phi[\lambda(\tau),E_{1}]$. That is
\begin{equation}
\begin{split}
\frac{1}{N!}\left[\frac{1}{h} \sqrt{2 \pi m} \frac{2}{a}\Gamma \left(1+\frac{1}{\lambda(0)}\right) \right]^{dN}  \frac{E_{0}^{\frac{dN}{2}+\frac{dN}{\lambda(0)}}}{\Gamma\left(\frac{dN}{2}+\frac{dN}{\lambda(0)}+1\right)}
=&\frac{1}{N!}\left[ \frac{1}{h} \sqrt{2 \pi m} \frac{2}{a}\Gamma \left(1+\frac{1}{\lambda(\tau)}\right) \right]^{dN}  \frac{E_{1}^{\frac{dN}{2}+\frac{dN}{\lambda(\tau)}}}{\Gamma\left(\frac{dN}{2}+\frac{dN}{\lambda(\tau)}+1\right)}\\
E_{1}^{\frac{dN}{2}+\frac{dN}{\lambda(\tau)}}=&\frac{\Gamma \left(1+\frac{1}{\lambda(0)}\right)^{dN}}{\Gamma \left(1+\frac{1}{\lambda(\tau)}\right)^{dN}} \frac{\Gamma\left(\frac{dN}{2}+\frac{dN}{\lambda(\tau)}+1\right)}{\Gamma\left(\frac{dN}{2}+\frac{dN}{\lambda(0)}+1\right)}E_{0}^{\frac{dN}{2}+\frac{dN}{\lambda(0)}}.
\end{split}
\end{equation}
or
\begin{equation}
E_{1}=\left[ \frac{\Gamma \left(1+\frac{1}{\lambda(0)}\right)}{\Gamma \left(1+\frac{1}{\lambda(\tau)}\right)}  \right]^{\frac{2\lambda(\tau)}{2+\lambda(\tau)}} \left[ \frac{\Gamma\left(\frac{dN}{2}+\frac{dN}{\lambda(\tau)}+1\right)}{\Gamma\left(\frac{dN}{2}+\frac{dN}{\lambda(0)}+1\right)} \right]^{\frac{2\lambda(\tau)}{dN[2+\lambda(\tau)]}} E_{0}^{\frac{[2+\lambda(0)]\lambda(\tau)}{\lambda(0)[2+\lambda(\tau)]}}.
\end{equation}
The Carnot cycle can be uniquely determined if the following parameters are fixed: Boltzmann constant $k_{B}=1$, at instant $B$, $\lambda_{B}=2$, and at instant $D$, $\lambda_{D}=4$. The temperatures of the two heat reservoirs are $T_{H}=4.6$, and $T_{C}=2.3$.

In the following we will try to determine $\lambda_{C}$ and $\lambda_{A}$.
The energy of the system at point $C'$
\begin{equation}
E_{1}=\left[ \frac{\Gamma \left(1+\frac{1}{\lambda_{B}}\right)}{\Gamma \left(1+\frac{1}{\lambda_{C}}\right)}  \right]^{\frac{2\lambda_{C}}{2+\lambda_{C}}} \left[ \frac{\Gamma\left(\frac{dN}{2}+\frac{dN}{\lambda_{C}}+1\right)}{\Gamma\left(\frac{dN}{2}+\frac{dN}{\lambda_{B}}+1\right)} \right]^{\frac{2\lambda_{C}}{dN(2+\lambda_{C})}} E_{0}^{\frac{[2+\lambda_{B}]\lambda_{C}}{\lambda_{B}[2+\lambda_{C}]}}.
\end{equation}
The probability distribution of energy at point $C'$
\begin{equation}
\begin{split}
P(E_{0}, \beta_{H})&=\frac{1}{Z_{N}}g(E_{0})e^{-\beta_{H} E_{0}}\\
&= \beta_{H} \frac{(\beta_{H}
E_{0})^{\frac{dN}{\lambda_{B}}+\frac{dN}{2}-1}}{\Gamma\left(\frac{dN}{\lambda_{B}}+\frac{dN}{2}\right)}
e^{-\beta_{H} E_{0}}. \label{probability}
\end{split}
\end{equation}
The average energy of the ensemble after the adiabatic evolution from $B$ to $C'$ is
\begin{equation}
\left\langle E_{C'} \right \rangle=\int_{0}^{\infty} dE_{0} P(E_{0}, \beta_{H}) \left[ \frac{\Gamma \left(1+\frac{1}{\lambda_{B}}\right)}{\Gamma \left(1+\frac{1}{\lambda_{C}}\right)}  \right]^{\frac{2\lambda_{C}}{2+\lambda_{C}}} \left[ \frac{\Gamma\left(\frac{dN}{2}+\frac{dN}{\lambda_{C}}+1\right)}{\Gamma\left(\frac{dN}{2}+\frac{dN}{\lambda_{B}}+1\right)} \right]^{\frac{2\lambda_{C}}{dN(2+\lambda_{C})}} E_{0}^{\frac{[2+\lambda_{B}]\lambda_{C}}{\lambda_{B}[2+\lambda_{C}]}}.
\end{equation}
By utilizing Eq~(\ref{probability}), we obtain
\begin{equation}
\left\langle E_{C'} \right \rangle=\left[ \frac{\Gamma \left(1+\frac{1}{\lambda_{B}}\right)}{\Gamma
\left(1+\frac{1}{\lambda_{C}}\right)}
\right]^{\frac{2\lambda_{C}}{2+\lambda_{C}}}\left[ \frac{\Gamma\left(\frac{dN}{2}+\frac{dN}{\lambda_{C}}+1\right)}{\Gamma\left(\frac{dN}{2}+\frac{dN}{\lambda_{B}}+1\right)} \right]^{\frac{2\lambda_{C}}{dN(2+\lambda_{C})}} \frac{\Gamma\left(\frac{dN}{2}+\frac{dN}{\lambda_{B}}+\frac{(2+\lambda_{B})\lambda_{C}}{\lambda_{B}(2+\lambda_{C})}\right)}{\Gamma\left(\frac{dN}{2}+\frac{dN}{\lambda_{B}}\right)} \beta_{H}^{-\frac{[2+\lambda_{B}]\lambda_{C}}{\lambda_{B}[2+\lambda_{C}]}}.
\end{equation}
The average energy of the ensemble after the relaxation is
\begin{equation}
\begin{split}
\left\langle E_{C} \right \rangle&=\int dE_{1} P(E_{1},\beta_{C}, \lambda_{C}) E_{1}\\
&=\int dE_{1} \beta_{C} \frac{(\beta_{C} E_{1})^{\frac{dN}{2}+\frac{dN}{\lambda_{C}}-1}}{\Gamma\left (\frac{dN}{2}+\frac{dN}{\lambda_{C}}\right)}
 \exp{[-\beta_{C}E_{1}]} E_{1}\\
&=\frac{1}{\beta_{C}}\left(\frac{dN}{2}+\frac{dN}{\lambda_{C}}
\right).
\end{split}
\end{equation}
The parameter $\lambda_{C}$ at point $C$ can be determined by the equation
$\left\langle E_{C'} \right \rangle=\left\langle E_{C} \right \rangle$. That is
\begin{equation}
\frac{1}{\beta_{C}}\left(\frac{dN}{2}+\frac{dN}{\lambda_{C}}
\right)=\left[ \frac{\Gamma \left(1+\frac{1}{\lambda_{B}}\right)}{\Gamma
\left(1+\frac{1}{\lambda_{C}}\right)}
\right]^{\frac{2\lambda_{C}}{2+\lambda_{C}}}\left[ \frac{\Gamma\left(\frac{dN}{2}+\frac{dN}{\lambda_{C}}+1\right)}{\Gamma\left(\frac{dN}{2}+\frac{dN}{\lambda_{B}}+1\right)} \right]^{\frac{2\lambda_{C}}{dN(2+\lambda_{C})}} \frac{\Gamma\left(\frac{dN}{2}+\frac{dN}{\lambda_{B}}+\frac{(2+\lambda_{B})\lambda_{C}}{\lambda_{B}(2+\lambda_{C})}\right)}{\Gamma\left(\frac{dN}{2}+\frac{dN}{\lambda_{B}}\right)} \beta_{H}^{-\frac{[2+\lambda_{B}]\lambda_{C}}{\lambda_{B}[2+\lambda_{C}]}}.
\label{temperature}
\end{equation}
Substituting $\lambda_{B}=2$, $T_{C}=2.3$, and $T_{H}=4.6$ into the above formula we obtain the value of $\lambda_{C}$. It is worth mentioning that $\lambda_{C}$ is a function of $N$, $\lambda_{C}=\lambda_{C}(N)$. We can also determine $\lambda_{A}$ at point
$A$ in a similar way. At point D, $\lambda_{D}=4$, we have the energy of the system at
point $A'$
\begin{equation}
\frac{1}{\beta_{H}}\left(\frac{dN}{2}+\frac{dN}{\lambda_{A}}
\right)=\left[ \frac{\Gamma \left(1+\frac{1}{\lambda_{D}}\right)}{\Gamma
\left(1+\frac{1}{\lambda_{A}}\right)}
\right]^{\frac{2\lambda_{A}}{2+\lambda_{A}}}\left[ \frac{\Gamma\left(\frac{dN}{2}+\frac{dN}{\lambda_{A}}+1\right)}{\Gamma\left(\frac{dN}{2}+\frac{dN}{\lambda_{D}}+1\right)} \right]^{\frac{2\lambda_{A}}{dN(2+\lambda_{A})}} \frac{\Gamma\left(\frac{dN}{2}+\frac{dN}{\lambda_{D}}+\frac{(2+\lambda_{D})\lambda_{A}}{\lambda_{D}(2+\lambda_{A})}\right)}{\Gamma\left(\frac{dN}{2}+\frac{dN}{\lambda_{D}}\right)} \beta_{C}^{-\frac{[2+\lambda_{D}]\lambda_{A}}{\lambda_{D}[2+\lambda_{A}]}}
\end{equation}
Through a similar procedure, we obtain the value of $\lambda_{A}$ at point $A'$. Thus the parameters of the Carnot cycle is completely
determined.

{\bf Relative entropy and the efficiency of the Carnot cycle:} In
order to calculate the efficiency as a function of the particle
number $N$, we need to obtain the thermodynamic entropy of the
system $S_{B}$, $S_{D}$ and the relative entropy
$D[\rho_{C'}||\rho^{eq}_{C}]=S_{C}-S_{C'}$ and
$D[\rho_{A'}||\rho^{eq}_{A}]=S_{A}-S_{A'}$. The entropy of the $N$
particle system at point $B$ is
\begin{equation}
\begin{split}
S_{B}&=-k_{B}\int_{0}^{\infty}dE_{0} g(E_{0}, \lambda_{B})
\left(\frac{\exp{[-\beta_{H}
E_{0}]}}{Z_{N}(\lambda_{B},\beta_{H})} \ln{\frac{\exp{[-\beta_{H} E_{0}]}}{Z_{N}(\lambda_{B},\beta_{H})}}\right)\\
&=\left(\frac{dN}{\lambda_{B}} +\frac{dN}{2}\right)k_{B}+dN k_{B}\ln
\left[\frac{1}{h}\frac{2}{a}\sqrt{2\pi m}\Gamma(1+\frac{1}{\lambda_{B}})
\right]-\left(\frac{dN}{\lambda_{B}} +\frac{dN}{2}\right)k_{B} \ln \beta_{H}-k_{B}\ln {N!}\\
&=\left(\frac{dN}{\lambda_{B}} +\frac{(d+2)N}{2}\right)k_{B}+N k_{B}\ln
\left[\frac{1}{N }\left(\frac{2}{ a}\right)^{d}\left(\frac{\sqrt{2\pi m}}{h}\right)^{d}\Gamma(1+\frac{1}{\lambda_{B}})^{d}
\right]-\left(\frac{dN}{\lambda_{B}} +\frac{dN}{2}\right)k_{B} \ln \beta_{H}.\\
\end{split}
\end{equation}
Similarly we can calculate the entropy $S_{D}$.
\begin{equation}
\begin{split}
S_{D}=\left(\frac{dN}{\lambda_{D}} +\frac{(d+2)N}{2}\right)k_{B}+N k_{B}\ln
\left[\frac{1}{N }\left(\frac{2}{ a}\right)^{d}\left(\frac{\sqrt{2\pi m}}{h}\right)^{d}\Gamma(1+\frac{1}{\lambda_{D}})^{d}
\right]-\left(\frac{dN}{\lambda_{D}} +\frac{dN}{2}\right)k_{B} \ln \beta_{C}.\\
\end{split}
\end{equation}
Hence we have
\begin{equation}
S_{B}-S_{D}=\left(\frac{dN}{\lambda_{B}} - \frac{dN}{\lambda_{D}}\right) k_{B} +dN k_{B}\ln{\frac{\Gamma(1+\frac{1}{\lambda_{B}})}{\Gamma(1+\frac{1}{\lambda_{D}})}}-\left(\frac{dN}{\lambda_{B}} +\frac{dN}{2}\right)k_{B} \ln \beta_{H}+\left(\frac{dN}{\lambda_{D}} +\frac{dN}{2}\right)k_{B} \ln \beta_{C}.
\end{equation}
This entropy is extensive, or proportional to $N$ when both $N$ and $2/a$ increase by the same ratio.
The entropy at point $C$ is
\begin{equation}
\begin{split}
S_{C}=\left(\frac{dN}{\lambda_{C}} +\frac{(d+2)N}{2}\right)k_{B}+N k_{B}\ln
\left[\frac{1}{N }\left(\frac{2}{ a}\right)^{d}\left(\frac{\sqrt{2\pi m}}{h}\right)^{d}\Gamma(1+\frac{1}{\lambda_{C}})^{d}
\right]-\left(\frac{dN}{\lambda_{C}} +\frac{dN}{2}\right)k_{B} \ln \beta_{C}.\\
\end{split}
\end{equation}
Because of the Liouville theorem, we have $S_{C'}=S_{B}$. Hence the entropy increase in the relaxation process can be expressed as
\begin{equation}
\begin{split}
D[\rho_{C'}||\rho^{eq}_{C}]=&S_{C}-S_{B}\\
=&\left(\frac{dN}{\lambda_{C}} -\frac{dN}{\lambda_{B}}\right)k_{B}+dN k_{B}\ln
\left[\frac{\Gamma(1+\frac{1}{\lambda_{C}})}{\Gamma(1+\frac{1}{\lambda_{B}})}
\right]-\left(\frac{dN}{\lambda_{C}} +\frac{dN}{2}\right)k_{B} \ln \beta_{C}+\left(\frac{dN}{\lambda_{B}} +\frac{dN}{2}\right)k_{B} \ln \beta_{H}.
\end{split}
\label{relative}
\end{equation}
From Eq. (\ref{temperature}) we have
\begin{equation}
\begin{split}
-\frac{2+\lambda_{C}}{2 \lambda_{C}} \ln \beta_{C} + \frac{2+\lambda_{C}}{2 \lambda_{C}} \ln \left[ \frac{dN}{2}+ \frac{dN}{\lambda_{C}} \right]&=\frac{2+\lambda_{C}}{2 \lambda_{C}}
\ln {\frac{\Gamma\left(\frac{dN}{2}+\frac{dN}{\lambda_{B}}+\frac{(2+\lambda_{B})\lambda_{C}}{\lambda_{B}(2+\lambda_{C})}\right)}{\Gamma\left(\frac{dN}{2}+\frac{dN}{\lambda_{B}}\right)} }\\
&+\frac{1}{dN}\ln{\left[ \frac{\Gamma\left(\frac{dN}{2}+\frac{dN}{\lambda_{C}}+1\right)}{\Gamma\left(\frac{dN}{2}+\frac{dN}{\lambda_{B}}+1\right)} \right]} + \ln{\left[ \frac{\Gamma \left(1+\frac{1}{\lambda_{B}}\right)}{\Gamma
\left(1+\frac{1}{\lambda_{C}}\right)}
\right]}-\frac{2+\lambda_{B}}{2 \lambda_{B}} \ln \beta_{H}.
\end{split}
\end{equation}
Substituting them into Eq. (\ref{relative}) we obtain
\begin{equation}
\begin{split}
D[\rho_{C'}||\rho^{eq}_{C}]=&\left(\frac{dN}{\lambda_{C}} -\frac{dN}{\lambda_{B}}\right)k_{B}+ k_{B}\ln{\left[ \frac{\Gamma\left(\frac{dN}{2}+\frac{dN}{\lambda_{C}}+1\right)}{\Gamma\left(\frac{dN}{2}+\frac{dN}{\lambda_{B}}+1\right)} \right]} \\
&+dNk_{B}\frac{2+\lambda_{C}}{2 \lambda_{C}}
\ln {\frac{\Gamma\left(\frac{dN}{2}+\frac{dN}{\lambda_{B}}+\frac{(2+\lambda_{B})\lambda_{C}}{\lambda_{B}(2+\lambda_{C})}\right)}{\Gamma\left(\frac{dN}{2}+\frac{dN}{\lambda_{B}}\right)} }
-dNk_{B} \frac{2+\lambda_{C}}{2 \lambda_{C}} \ln \left[ \frac{dN}{2}+ \frac{dN}{\lambda_{C}} \right]. \\
\end{split}
\end{equation}
Interestingly this relative entropy is independent of the temperatures. The relative entropy $D[\rho_{C'}||\rho^{eq}_{C}]$ can be obtained when $\lambda_{B}$, and $\lambda_{C}$ are fixed. If we choose $d=3$ the relative entropy can be further simplified to
\begin{equation}
\begin{split}
D[\rho_{C'}||\rho^{eq}_{C}]=&3Nk_{B}\frac{2-\lambda_{C}}{2 \lambda_{C}}+ k_{B}\ln{\left[ \frac{\Gamma\left(\frac{3N}{2}+\frac{3N}{\lambda_{C}}+1\right)}{\Gamma\left(3N+1\right)} \right]} \\
&+3Nk_{B}\frac{2+\lambda_{C}}{2 \lambda_{C}}
\left\{\ln {\frac{\Gamma\left(3N+\frac{2\lambda_{C}}{(2+\lambda_{C})}\right)}{\Gamma\left(3N\right)} }
- \ln \left[ \frac{3N}{2}+ \frac{3N}{\lambda_{C}} \right] \right\},\\
\end{split}
\end{equation}
where $\lambda_{C}(N)$ is determined by Eq.~(\ref{temperature}).
One can further simplify the relative entropy when taking the limit of $N\to \infty$
\begin{equation}
\lim \limits_{N \to \infty} D[\rho_{C'}||\rho^{eq}_{C}]\approx k_{B} \frac{2 \lambda_{C}}{2+\lambda_{C}}.
\end{equation}
The relative entropy $D[\rho_{A'}||\rho^{eq}_{A}]$ can also be obtained in a similar way when $\lambda_{D}$ and $\lambda_{A}$ are fixed.



\begin{thebibliography}{99}
\bibitem{explain} The main result (\ref{efficiency})
is valid even when the adiabatic process is not quasistatic, because one can prove that in a non-quasistatic process $S_{l}-S_{l'}=D[\rho_{l'} || \rho_{l}^{\mathrm{eq}}  ]$ also holds true as long as no heat exchange between the system and the reservoir occurs in the relaxation processes $C \to C'$ and $A \to A'$. This can be proved straightforwardly via the relation $F_{l}-F_{l'}=D[  \rho_{l'}|| \rho_{l}^{\mathrm{eq}}]$ \cite{relativeentropy1}. Nevertheless, if the adiabatic process is not quasistatic, the efficiency (\ref{efficiency}) is not the maximum efficiency, and in the thermodynamic limit $N\to\infty$ it will not approach the Carnot effiency. In the current study we focus on the
quasistatic process only.
\end{thebibliography}
\end{document}